\documentclass[twocolumn,prc,superscriptaddress,unsortedaddress,showpacs,preprintnumbers,amsmath,amssymb]{revtex4}

\usepackage{graphicx}
\usepackage{dcolumn}
\usepackage{bm}
\usepackage{CJK}
\usepackage{CJKulem}
\usepackage{txfonts}

\def\beq{\begin{equation}}
\def\eeq{\end{equation}}
\def\bea{\begin{eqnarray}}
\def\eea{\end{eqnarray}}

\def\fun#1#2{\lower3.6pt\vbox{\baselineskip0pt\lineskip.9pt
  \ialign{$\mathsurround=0pt#1\hfil##\hfil$\crcr#2\crcr\sim\crcr}}}

\begin{document}
\begin{CJK*} {GBK} {song}
\preprint{}

\title{Origin of symmetry energy in finite nuclei and density dependence of nuclear matter
symmetry energy from measured alpha-decay energies}

\author{Jianmin Dong}\affiliation{Institute of Modern Physics, Chinese
Academy of Sciences, Lanzhou 730000, China}
\author{Wei Zuo}\email[ ]{zuowei@impcas.ac.cn}
\affiliation{Institute of Modern Physics, Chinese Academy of
Sciences, Lanzhou 730000, China}
\author{Jianzhong Gu}
\affiliation{China Institute of Atomic Energy, P. O. Box 275(10),
Beijing 102413, China}

\date{\today}

\begin{abstract}
Based on the Skyrme energy density functional, the spatial
distribution of the symmetry energy of a finite nucleus is derived
in order to examine whether the symmetry energy of a finite nucleus
originates from its interior or from its surface. It is found that
the surface part of a heavy nucleus contributes dominantly to its
symmetry energy compared to its inner part. The symmetry energy
coefficient $a_{\text{sym}}({A})$ is then directly extracted and the
ratio of the surface symmetry coefficient to the volume symmetry
coefficient $\kappa$ is estimated. Meanwhile, with the help of
experimental alpha decay energies, a macroscopic method is developed
to determine the symmetry energy coefficient of heavy nuclei. The
resultant $a_{\text{sym}}({A})$ is used to analyze the density
dependence of the symmetry energy coefficient of nuclear matter
around the saturation density, and furthermore, the neutron skin
thickness of $^{208}\text{Pb}$ is deduced which is consistent with
the pygmy dipole resonance analysis. In addition, it is shown that
the ratio $\kappa$ obtained from the macroscopic method is in
agreement with that from the Skyrme energy density functional. Thus
the two completely different approaches may validate each other to
achieve more compelling results.

\end{abstract}


\maketitle
\section{Introduction}\label{intro}\noindent
Great attention has been paid to the nuclear equation of state (EOS)
of isospin asymmetric nuclear matter, in particular the nuclear
matter symmetry energy coefficient $S(\rho )$. The density-dependent
symmetry energy coefficient plays a crucial role in understanding a
variety of issues in nuclear physics as well as astrophysics, such
as the heavy ion reactions \cite{PD,AWS,VB,BAL,JML}, the stability
of superheavy nuclei \cite{JD}, the structures, composition and
cooling of neutron stars \cite{NS1,BKSP,NS,BG} and even some new
physics beyond the standard model \cite{B1,B2}. The energy per
particle in nuclear matter is $e(\rho ,\delta )=e(\rho ,0)+S(\rho
)\delta ^{2}+\mathcal{O(}\delta ^{4})$ with the density $\rho =\rho
_{n}+\rho _{p}$ and isospin asymmetry $\delta =(\rho _{n}-\rho
_{p})/\rho $. The density-dependent symmetry energy coefficient
$S(\rho )$ can be expanded to second order by $S(\rho
)=S_{0}+L\epsilon +K_{\text{sym}}\epsilon ^{2}/2$ with $\epsilon
=(\rho -\rho _{0})/3\rho _{0}$, where $S_{0}$ is the symmetry energy
coefficient at the nuclear matter saturation density, $L=3\rho
\partial S(\rho )/\partial \rho |_{\rho _{0}}$ and
$K_{\text{sym}}=9\rho ^{2}\partial ^{2}S/\partial \rho ^{2}|_{\rho
_{0}}$ are the slope and curvature parameters governing the density
dependence of $S(\rho )$ around the saturation density $\rho_{0}$.

Various independent investigations have been carried out to
constrain the density dependence of the nuclear matter symmetry
energy coefficient. A microscopic Brueckner-Hartree-Fock approach
using the realistic Argonne V18 nucleon-nucleon potential plus a
phenomenological three-body force gives a value of the slope
parameter $L = 66.5$ MeV \cite{AV18}. In a description of isospin
diffusion data using a Boltzmann-Uehling-Uhlenbeck (BUU) transport
model (IBUU04) with free-space experimental nucleon-nucleon (NN)
cross section, Chen {\it et al.} extracted a nuclear symmetry energy
coefficient of $S(\rho )\approx S_{0}(\rho /\rho _{0})^{\gamma}$
with $\gamma=1.05$ by comparing the theoretical results with the
experimental data \cite{LWC}. Including the medium-dependent NN
cross section, the isospin diffusion data lead to a softer symmetry
energy coefficient with $\gamma=0.69$. The pygmy dipole resonance
(PDR) of $^{208}$Pb analyzed with Skyrme interactions gives
$L=64.8\pm15.7$ MeV \cite{PDR}. Liu {\it et al.} extracted the
symmetry energy coefficients for the finite nuclei with mass number
$A=20-250$ from more than 2000 measured nuclear masses \cite{ML}.
With the semiempirical relationship between the symmetry energy
coefficients $a_{\text{sym}}({A})$ of finite nuclei and $S(\rho )$
of nuclear matter, they obtained the slope parameter $L=53-79$ MeV
at the normal density. The analysis of isospin diffusion and double
ratio data involving neutron and proton spectra by an improved
quantum molecular dynamics transport model suggests $S(\rho
)=12.5\left( \rho /\rho _{0}\right) ^{2/3}+C_{p}\left( \rho /\rho
_{0}\right) ^{\gamma }$ \cite{MBT}with $\gamma =0.4-1.05$.
Danielewicz and Lee obtained the $L$ values ranging from 78 to 111
MeV \cite{DL}. Centelles {\it et al.} recently reported the values
of $L=75\pm25$ MeV and later revised to a narrow window of $L=45-75$
from neutron skin thickness of nuclei \cite{MC,MC2}. With a
correlation for the symmetry energy at the saturation density
$S_{0}$, the slope parameter $L$ and the curvature parameter
$K_{\text{sym}}$ based on widely different mean-field interactions,
Ref. \cite{OUR} suggested that the $L$ value is $56\pm24$ MeV. The
neutron skin thickness $\Delta R_{np}$ given by the difference of
neutron and proton root-mean-square radii, correlates linearly with
the slope parameter $L$ \cite{BAB,ST,RJF}. Therefore, a measurement
of $\Delta R_{np}$ with a high accuracy should be a strong
constraint on the density dependence of $S(\rho )$. Nevertheless,
the density dependence of $S(\rho )$ at subnormal density is
currently still an open question, and further investigations are
required.

Recently, the symmetry energy of finite nuclei has been widely
investigated because with the help of it one may gain some
information on the density dependence of $S(\rho )$. The most
commonly used methods are the liquid drop models combined
with nuclear masses and the leptodermous expansion based on
the self-consistent mean-field theory \cite{ML,AA1,AA2,AA3}.
In this study, we extract directly the symmetry energy coefficient
of finite nuclei $a_{\text{sym}}({A})$ by using the Skyrme energy
density functional and mainly focus on the spatial distribution
of the symmetry energy of a heavy nucleus. We would point out that
compared to previous studies, we employ a different approach that
the leptodermous expansion is not applied to a mean-field theory.
Another purpose of the present work is to extract $a_{\text{sym}}({A})$
based on experimental alpha decay energies of heavy nuclei and investigate
the density dependence of $S(\rho )$.

This work is organized as follows. In Sec. II, the approach to
construct the symmetry energy density functional of finite nuclei is
presented. $a_{\text{sym}}({A})$ and its spatial distribution is
calculated and analyzed. In Sec. III, $a_{\text{sym}}({A})$ is
extracted by using the experimental alpha-decay energies and the
density dependence of $S(\rho )$ is then determined. Finally a
summary is given in Sec. IV.

\section{Symmetry energy coefficient of finite nuclei from the Skyrme energy density functional}\label{intro}\noindent
The Skryme interaction has been a powerful tool in the investigation
of finite nuclei and nuclear matter \cite{SHF8,SHF9,SHF10,SHF11}. In
the standard Skyrme-Hartree-Fock (SHF) method, the interaction is a
zero-range, density- and momentum-dependent form. The expression for
the total energy density functional is
\begin{eqnarray}
\mathcal{H} &\mathcal{=}&\frac{1}{2}\hbar ^{2}\left( f_{p}\tau
_{p}+f_{n}\tau _{n}\right)   \notag \\
&&+\left[ \frac{t_{0}}{2}\left( 1+\frac{x_{0}}{2}\right) +\frac{t_{3}}{12}%
\left( 1+\frac{x_{3}}{2}\right) \rho ^{\alpha }\right] \rho ^{2}  \notag \\
&&-\left[ \frac{t_{0}}{2}\left( x_{0}+\frac{1}{2}\right) +\frac{t_{3}}{12}%
\left( x_{3}+\frac{1}{2}\right) \rho ^{\alpha }\right] \left( \rho
_{n}^{2}+\rho _{p}^{2}\right)   \notag \\
&&+\left[ \frac{3t_{1}}{16}\left( 1+\frac{x_{1}}{2}\right) -\frac{t_{2}}{16}%
\left( 1+\frac{x_{2}}{2}\right) \right] \left( \nabla \rho \right)
^{2}
\notag \\
&&-\left[ \frac{3t_{1}}{16}\left( x_{1}+\frac{1}{2}\right) +\frac{t_{2}}{16}%
\left( x_{2}+\frac{1}{2}\right) \right] \left( \left( \nabla \rho
_{n}\right) ^{2}+\left( \nabla \rho _{p}\right) ^{2}\right)   \notag \\
&&+\frac{1}{16}(t_{1}-t_{2})\left( J_{n}^{2}+J_{p}^{2}\right) -\frac{1}{16}%
(t_{1}x_{1}+t_{2}x_{2})J^{2}  \notag \\
&&+\frac{1}{2}w_{0}\left[ \overrightarrow{J}\cdot \nabla \rho +%
\overrightarrow{J_{n}}\cdot \nabla \rho
_{n}+\overrightarrow{J_{p}}\cdot
\nabla \rho _{p}\right]   \notag \\
&&+\frac{e^{2}}{2}\rho _{p}\int \frac{\rho _{p}(r^{\prime
})}{\left\vert
\mathbf{r}-\mathbf{r}^{\prime }\right\vert }d\mathbf{r}^{\prime }-\frac{%
3e^{2}}{4}\left( \frac{3}{\pi }\right) ^{1/3}\rho _{p}^{4/3},
\end{eqnarray}
where $f_{n}$ and $f_{p}$ are given by
\begin{eqnarray}
f_{p} &=&\frac{1}{m}+\frac{1}{2\hbar ^{2}}\left[ t_{1}\left( 1+\frac{x_{1}}{2%
}\right) +t_{2}\left( 1+\frac{x_{2}}{2}\right) \right] \rho \nonumber \\
&&-\frac{1}{2\hbar ^{2}}\left[ t_{1}\left( x_{1}+\frac{1}{2}\right)
-t_{2}\left( x_{2}+\frac{1}{2}\right) \right] \rho _{p},
\end{eqnarray}%
\begin{eqnarray}
f_{n} &=&\frac{1}{m}+\frac{1}{2\hbar ^{2}}\left[ t_{1}\left( 1+\frac{x_{1}}{2%
}\right) +t_{2}\left( 1+\frac{x_{2}}{2}\right) \right] \rho  \nonumber\\
&&-\frac{1}{2\hbar ^{2}}\left[ t_{1}\left( x_{1}+\frac{1}{2}\right)
-t_{2}\left( x_{2}+\frac{1}{2}\right) \right] \rho _{n}.
\end{eqnarray}
The $t_{0}$-$t_{3}$, $x_{0}$-$x_{3}$ and $\alpha$ are Skyrme
interaction parameters resulting from a fit to the binding energies
and charge radii of a number of known nuclei. $\rho _{q},\tau _{q}$
and $\overrightarrow{J}_{q}$ are the local nucleon density, kinetic
energy density and spin density for neutron or proton respectively,
whereas $\rho ,\tau $ and $\overrightarrow{J}$ are the corresponding
total densities. After a self-consistent SHF calculation, one
obtains the proton and neutron distributions $\rho _{p}(r)$ and
$\rho _{n}(r)$ in a nucleus. By using the obtained SHF density
distributions, the total energy of the nucleus can be calculated. In
order to avoid the implicit use of single particle wave functions
for calculating the contributions of the kinetic energy to the
symmetry energy, we express the kinetic energy density as a function
of the density $\rho _{q}(r)$ by using the extended-Thomas-Fermi
approximation to second order terms for reasons of being enough for
numerical convergence \cite{ETF}, and it is written as (excluding
the spin part) \cite{ETF,ETF1,ETF2}
\begin{eqnarray}
\tau _{q}^{(0)} &=&\frac{3}{5}\left( 3\pi ^{2}\right) ^{2/3}\rho
_{q}^{5/3},
\notag \\
\tau _{q}^{(2)} &=&\frac{1}{3}\nabla ^{2}\rho _{q}+\frac{\left(
\nabla \rho
_{q}\right) ^{2}}{36\rho _{q}}+\frac{\nabla f_{q}\cdot \nabla \rho _{q}}{%
6f_{q}}-  \notag \\
&&\frac{\rho _{q}\left( \nabla f_{q}\right)
^{2}}{12f_{q}^{2}}+\frac{\rho _{q}\nabla ^{2}f_{q}}{6f_{q}},
\end{eqnarray}
where $q$ denotes neutron or proton. Here, we stress that the
extend-Thomas-Fermi approximation is only adopted for calculating
the total kinetic energy after the self-consistent SHF solution. Let
$\rho _{n}=\rho _{p}=\rho/2$ in the above formula (symmetry case),
the density functional is given by $\mathcal{H}_{0}$. Therefore,
excluding the Coulomb energy and the spin energy in
$\mathcal{H}-\mathcal{H}_{0}$, the density functional for the
symmetry energy is written as
\begin{equation}
\mathcal{H}_{\text{sym}}=\mathcal{H}_{T}+%
\mathcal{H}_{V}+\mathcal{H}_{grad},
\end{equation}
with
\begin{equation}
\mathcal{H}_{T}=\frac{\hbar ^{2}}{2}\left( f_{p}\tau _{p}+f_{n}\tau
_{n}-2f_{n=p}\tau _{n=p}\right),
\end{equation}
\begin{equation}
\mathcal{H}_{V}=-\left[ \frac{t_{0}}{2}\left( x_{0}+\frac{1}{2}\right) +%
\frac{t_{3}}{12}\left( x_{3}+\frac{1}{2}\right) \rho ^{\alpha }\right] \frac{%
1}{2}\rho ^{2}\delta ^{2},
\end{equation}
\begin{equation}
\mathcal{H}_{grad}=-\left[ \frac{3t_{1}}{16}\left( x_{1}+\frac{1}{2}\right) +%
\frac{t_{2}}{16}\left( x_{2}+\frac{1}{2}\right) \right]
\frac{1}{2}[\nabla (\rho \delta )]^{2}.
\end{equation}
In a self-consistent SHF calculation, one obtains the nucleon
density distribution finally. With the nucleon density
distributions, the $\mathcal{H}_{T}$, $\mathcal{H}_{V}$ and
$\mathcal{H}_{grad}$ can be estimated accordingly. The total
symmetry energy of a finite nucleus is defined as
$E_{\text{sym}}(A)=a_{\text{sym}}(A)(N-Z)^{2}/A=\int_{0}^{\infty }\mathcal{H}_{%
\text{sym}}dV$. Here the symmetry energy coefficient
$a_{\text{sym}}(A)$ includes the volume and surface symmetry energy
coefficient.

\begin{figure}[htbp]
\begin{center}
\includegraphics[width=0.5\textwidth]{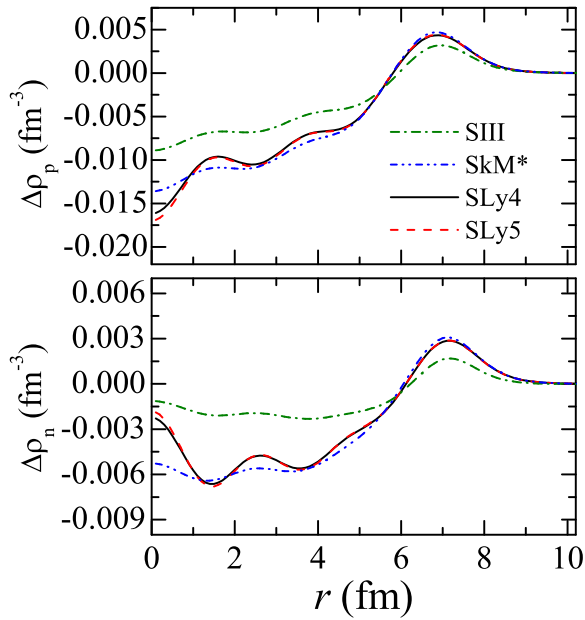}
\caption{(Color Online) Polarizations of the neutron and proton
densities of $^{208}$Pb induced by Coulomb interaction.}
\end{center}
\end{figure}

\begin{table}[h]
\label{table1} \caption{The calculated symmetry energy coefficient
$a_{\text{sym}}$ for $^{208}$Pb. The contributions of the gradient
term $a_{\text{grad}}=\int_{0}^{\infty }4\pi
\mathcal{H}_{\text{grad}}r^{2}dr$ and the second order term of the
kinetic part $a_{\text{kin2}}=\int_{0}^{\infty }4\pi
r^{2}\mathcal{H}_{T}^{(2)}dr$ as well as their proportions in
$a_{\text{sym}}$ are all listed. $\kappa$ is the ratio of the
surface symmetry coefficient to the volume symmetry coefficient
$S_{0}$.}

\begin{ruledtabular}
\begin{tabular}{llllllllllllllll}
  & SIII  & SLy4 & SLy5 & SkM*  \\
\hline
$a_{\text{sym}}$ (MeV) & 21.6 & 23.6 & 23.3 & 20.8 \\

$a_{\text{grad}}$ (MeV) & -0.290 & -0.348 & -0.399 & -0.307 \\

$a_{\text{grad}}/a_{\text{sym}}$ & $-1.3\%$ & $-1.5\%$ & $-1.7\%$ &
$-1.5\%$ \\

$a_{\text{kin2}}$ (MeV)& 0.365 & 0.543 & 0.516 & 0.136 \\

$a_{\text{kin2}}/a_{\text{sym}}$ &$1.7\%$ &$2.3\%$ &$2.2\%$
&$0.6\%$ \\
$S_{0}$ (MeV) & 28.2 & 32.0 & 32.0& 30.0 \\
$\kappa$ & 1.81 & 2.26 & 2.21 &2.62 \\

\end{tabular}
\end{ruledtabular}
\end{table}

Indeed, in order to extract the nuclear symmetry energy that relates
solely to the nuclear force, one should subtract the effect due to
the fact that the Coulomb interaction effectively polarizes the
neutron and proton densities. One can calculate the neutron or
proton density $\rho$ by using the Skyrme interactions, and when the
Coulomb interaction is switched off one obtains the density
$\rho^{'}$. The differences between $\rho$ and $\rho^{'}$, namely
$\Delta\rho_{n}=\rho_{n}-\rho_{n}^{'}$ and
$\Delta\rho_{p}=\rho_{p}-\rho_{p}^{'}$, denote the Coulomb
polarization effect. Fig. 1 displays the polarizations of the proton
(upper panel) and neutron (lower panel) densities of $^{208}$Pb
induced by the Coulomb interaction. The interactions used here (see
in Fig.1) yield the binding energy for $^{208}$Pb being 1636.62 MeV,
1636.47 MeV, 1635.67 MeV and 1633.58 MeV, respectively, which are in
very good agreement with the experimental measurement of 1635.84 MeV
\cite{mass}. The protons and neutrons move slightly away from the
core as a result of the Coulomb repulsion between charged protons.
For heavy nuclei such as $^{208}$Pb, this effect is weak while for
the nuclei with $N\approx Z$ this effect can be very
distinguishable.

We calculate the symmetry energy of $^{208}$Pb where the Coulomb
polarization effect of the nucleon densities is subtracted. The
calculated results are listed in Table I. The radial symmetry energy
density distributions $g(r)=4\pi\mathcal{H}_{\text{sym}}r^{2}$ as a
function of distance $r$ are displayed in Fig. 2. Different
interactions provide almost the same shape of the symmetry energy
density distribution. The inset in Fig. 2 shows the nucleon density
distribution in $^{208}$Pb, from which one can see that the nuclear
surface region is $6-8$ fm away from the nuclear centre. Though the
matter in this region is much less dense than that in the nuclear
interior, the contribution from the nuclear surface region to
$E_{\text{sym}}({A})$ is rather large. The reason is twofold:
Firstly, the spherical shell of the surface region has a large
volume. Secondly, the asymmetry distribution function
$\delta(r)=(\rho _{n}(r)-\rho _{p}(r))/\rho(r)$ has large values at
the surface region. Fig. 3 shows the isospin asymmetry function
$\delta(r)$ versus distance $r$. $\delta(r)$ is low in the interior
of the nucleus but increases rapidly at the surface region. This
also implies a neutron skin in $^{208}$Pb. The region of $r<3$ fm
contributes a little to $E_{\text{sym}}({A})$ since it has only a
small volume and $\delta(r)$ is low in this region in spite of its
high density. That the nuclear surface contributes greatly to
$a_{\text{sym}}({A})$ indicate that the symmetry energy coefficient
of a finite nucleus is considerably smaller than that of nuclear
matter at saturation density due to the low density at the surface.

\begin{figure}[htbp]
\begin{center}
\includegraphics[width=0.5\textwidth]{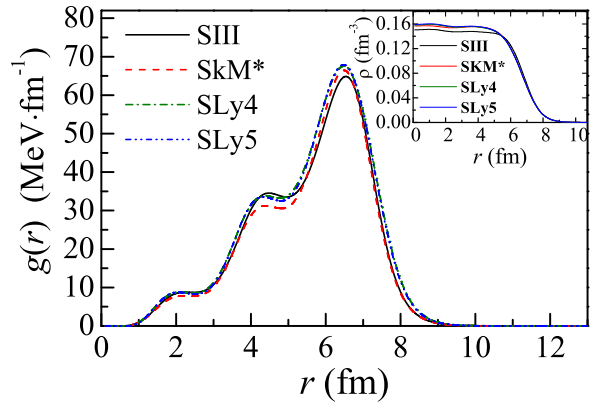}
\caption{(Color Online) Symmetry energy density distribution
$g(r)=4\pi \mathcal{H}_{\text{sym}}r^{2}$ as a function of $r$ in
$^{208}$Pb.}
\end{center}
\end{figure}

\begin{figure}[htbp]
\begin{center}
\includegraphics[width=0.5\textwidth]{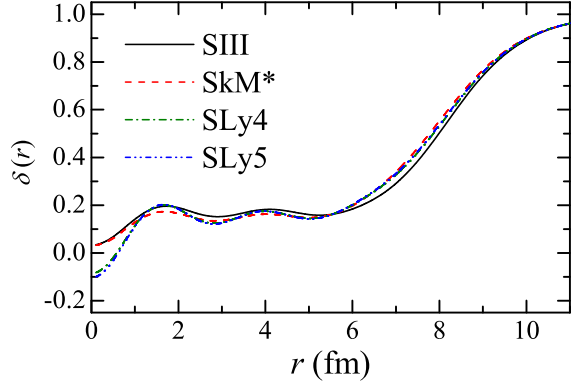}
\caption{(Color Online) Isospin asymmetry distribution function
$\delta(r) =(\rho _{n}(r)-\rho _{p}(r))/\rho(r) $ as a function of
distance $r$ in $^{208}$Pb.}
\end{center}
\end{figure}

The reduction of the $a_{\text{sym}}$ for $^{208}$Pb caused by the gradient term is
less than $2\%$, as presented in Table I. Also, the second order
term of the kinetic energy part contributes rather weakly to
the $a_{\text{sym}}$ (less than $3\%$). Moreover, the contributions from
these two weak terms tend to cancel out to a large extent. Thus one
can neglect the two terms simultaneously without loss of accuracy.
Accordingly, the $\mathcal{H}_{T}$ for finite nuclei may be
simplified as
\begin{eqnarray}
\mathcal{H}_{T} &=&\frac{2}{3}\left( 3\pi ^{2}\right) ^{2/3}\bigg\{\left( \frac{%
\rho }{2}\right) ^{5/3}+  \notag \\
&&\frac{m}{\hbar ^{2}}\left[ t_{1}(1+\frac{x_{1}}{2})+t_{2}(1+\frac{x_{2}}{2}%
)\right] \left( \frac{\rho }{2}\right) ^{8/3}+  \notag \\
&&\frac{2m}{\hbar ^{2}}\left[ t_{2}(x_{2}+\frac{1}{2})-t_{1}(x_{1}+\frac{1}{2%
})\right] \left( \frac{\rho }{2}\right) ^{8/3}\bigg\}\delta ^{2},
\end{eqnarray}
which leads to $\mathcal{H}_{sym}=\mathcal{H}_{T}+\mathcal{H}_{V}$
just a function of $\rho(r)$ and $\delta(r)$. For the large distance
$r$, the asymmetry $\delta (r)$ is close to one, but this region
almost does not contribute to the symmetry energy. Accordingly, this
equation of the $\mathcal{H}_{T}$ is valid.

The mass dependence of the symmetry energy coefficients of nuclei is
written as \cite{DROP,DL}
\begin{equation}
a_{\text{sym}}(A)=\frac{S_{0}}{1+\kappa A^{-1/3}},\text{with }\kappa =\frac{9%
}{4}\frac{S_{0}}{Q}, \label{F}
\end{equation}
where $\kappa$ is the ratio of the surface symmetry coefficient to
the volume symmetry coefficient and $Q$ is the surface stiffness
that measures the resistance of a nucleus against separation of
neutrons from protons to form a neutron skin. Since both the
$a_{\text{sym}}(A)$ and $S_{0}$ can be estimated within the Skyrme
interactions, the relation between the surface symmetry energy and
the volume symmetry energy characterized by the $\kappa$ value can
be investigated directly. The $\kappa$ value of $^{208}$Pb we
obtained is $1.8-2.6$, which agrees with that from the nuclear
masses \cite{ML} but slightly larger than that from the leptodermous
expansion based on the self-consistent mean-field theory \cite{AA1}
and from the binding energies in Ref. \cite{AA2}. Recently,
Centelles {\it et al}. proposed that the $a_{\text{sym}}(A)$ of
finite nuclei is approximately equal to $S(\rho _{A})$ of the
nuclear matter at a reference density $\rho_{A}$, namely
$S(\rho_{A})=a_{\text{sym}}({A})$ \cite{MC}. For example, the
reference density for $^{208}$Pb is $\rho _{A}=0.1$ fm$^{-3}$,
namely, $S(\rho _{A}=0.1
\text{fm}^{-3})=a_{\text{sym}}(^{208}\text{Pb})$. This relation
links the properties of finite nuclei and nuclear matter and helps
us to explore the density dependence of the nuclear symmetry energy
coefficient $S(\rho )$. The value of $S(\rho)$ at $\rho=0.1 \text{
fm}^{-3}$ has been widely investigated recently. The analysis of the
GDR of $^{208}$Pb with Skyrme interactions suggests $S(\rho =0.10$
fm$^{-3})=23.3-24.9$ MeV \cite{GDR} while the analysis of the GDR of
$^{132}$Sn within relativistic mean field models leads to $S(\rho
=0.10$ fm$^{-3})=21.2-22.5$ MeV \cite{LGC}. Our calculated
$a_{\text{sym}}(A)$ for $^{208}$Pb is 20.8--23.6 MeV which agrees
with that in Ref. \cite{ML} ($20.22-24.74$ MeV) obtained from the
measured nuclear masses, indicating the reliability of the present
approach to some extent.

\section{Density dependence of the nuclear matter symmetry energy from experimental Alpha-decay energies}\label{intro}\noindent
The nuclear symmetry energy coefficient $a_{\text{sym}}(A)$ is
usually extracted by directly fitting the measured nuclear masses
with different versions of the liquid-drop models
\cite{ML,WANG,TD0}. In this study, however, we extract
$a_{\text{sym}}({A})$ with the experimental alpha decay energies of
heavy (or superheavy) nuclei. In our previous work, a formula for
$Q_{\alpha}$ values was put forward which works well for heavy
nuclei with $Z\geq 92$ and $N\geq140$ \cite{DONG}. That formula will
be used here to extract $a_{\text{sym}}(A)$. And we use 162 heavy
and superheavy nuclei in the extraction. The present approach is
simple and physically clear since some less important terms in the
mass formula of the liquid drop model are canceled out to a great
extent in the calculation of $Q_{\alpha}$ values. In this mass
region, the shell correction energy is very small compared to the
other terms \cite{DONG}. So one is free from the influence of the
shell effects. Subtracting the Coulomb term $Q_{C}$, one obtains
\begin{eqnarray}
Q_{\alpha }-Q_{C}=-\frac{4S_{0}}{1+\kappa A^{-1/3}}\left(
\frac{N-Z}{A}\right) ^{2}+e,
\end{eqnarray}
where $Q_{C}=aZA^{-4/3}(3A-Z)$, $a=0.9373$ and $e$ is a constant.
The mass dependence of the symmetry energy coefficient is given by
Eq. (\ref{F}).

To fit $S_{0}$ and $\kappa$ directly is difficult to yield their
optimal values since many different combinations of $S_{0}$ and
$\kappa$ could provide the same least deviation. Considering that
$S_{0}$ has been relatively well determined as $31.6^{+2.2}_{-2.2}$
MeV nowadays \cite{OUR}, we solely determine the value of $\kappa$
in terms of $a_{\text{sym}}(A)$. The calculated value of $\kappa$ is
$2.21^{+0.56}_{-0.55}$ with all the root-mean-square deviations of
0.33 MeV for $Q_{\alpha }-Q_{C}$, where the uncertainty mainly
results from the uncertainty of $S_{0}$ value. The $\kappa$ value
here from the experimental $\alpha$-decay energies covers the range
given by the Skyrme energy density functional method (see Table I).
The two completely different approaches provide the unanimous
results and hence they may validate each other to get more
compelling results.

Now let us constrain the density dependence of the nuclear matter
symmetry energy coefficient at subnormal densities with the help of
the obtained $a_{\text{sym}}(^{208}\text{Pb})$. We use the Eq. (6)
in Ref. \cite{OUR} to describe $S(\rho)$, which reads
\begin{equation}
S(\rho )=17.47\left( \frac{\rho }{\rho _{0}}\right) ^{2/3}+C_{1}\left( \frac{%
\rho }{\rho _{0}}\right) +C_{2}\left( \frac{\rho }{\rho _{0}}\right)
^{1.52}, \label{X}
\end{equation}
where $C_{1}$ and $C_{2}$ are linked by $S_{0}=17.47+C_{1}+C_{2}$
and they need to be determined. We would point out that Eq.
(\ref{X}) is more universal for the description of the density
dependence of $S(\rho)$ \cite{OUR} compared with other forms. With
$S(\rho _{A})=a_{\text{sym}}({A})$ and Eq. (\ref{X}), we obtain
\begin{equation}
\frac{S_{0}}{1+\kappa A^{-1/3}}=17.47\left( \frac{\rho _{A}}{\rho
_{0}}\right)
^{2/3}+C_{1}\left( \frac{\rho _{A}}{\rho _{0}}\right) +C_{2}\left( \frac{%
\rho _{A}}{\rho _{0}}\right) ^{1.52}.\label{G}
\end{equation}
Inserting the detailed values of $S_{0}= 31.6^{+2.2}_{-2.2}$ MeV,
$\rho _{A}=0.10$ fm$^{-3}$ for $^{208}$Pb and
$\kappa=2.21^{+0.56}_{-0.55}$, we obtain
$C_{1}=24.56^{-7.82}_{+7.58}$ MeV and
$C_{2}=-10.43^{+10.02}_{-9.78}$ MeV, and accordingly the slope and
curvature parameters are calculated to be $L=61^{+22}_{-22}$ MeV and
$K_{\text{sym}}=-109^{+71}_{-70}$ MeV.

\begin{table}[h]
\label{table2} \caption{Values of the neutron skin thickness in
$^{208}$Pb obtained from various independent studies.}
\begin{ruledtabular}
\begin{tabular}{llllllllllllllll}
Reference  & Method  & $\Delta R_{np}$ (fm)  \\
\hline
Ref. \cite{PDR} & pygmy dipole resonance &  $0.194\pm0.024$ \\

Ref. \cite{AK} & pygmy dipole resonance &   $0.180\pm0.035$ \\

Ref. \cite{EDR}  & proton inelastic scattering &  $0.156^{+0.025}_{-0.021}$   \\

Ref. \cite{JZ}  &proton elastic scattering &  $0.211^{+0.054}_{-0.063}$   \\

Ref. \cite{KJC}  &chiral effective field theory &  $0.17\pm0.03$   \\

Ref. \cite{EA}  &exotic atom &  $0.18\pm0.02$   \\

Ref. \cite{JBG}  & electric dipole polarizability &  $0.168\pm0.022$   \\

Ref. \cite{OUR}  &mean-field  &  $0.185\pm0.035$   \\

Present & alpha decay energies &   $0.191\pm0.032$  \\
\end{tabular}
\end{ruledtabular}
\end{table}

It has been well established in Ref. \cite{XRM} that the correlation
between $L$ and $\Delta R_{np}$ is $\Delta R_{np}=0.101 + 0.00147L$
for $^{208}$Pb, and here $L$ and $\Delta R_{np}$ are measured in
units of fm. With the values of $L$ we obtained, the neutron skin
thickness in $^{208}$Pb is $\Delta R_{np}=0.191\pm0.032$ fm in terms
of this correlation. A comparison of our result with that of others
is presented in Table II. On the whole, our result is in agreement
with the others, especially the very recent data obtained from the
pygmy dipole resonance. However, the values of the neutron skin
thickness in $^{208}$Pb obtained by various approaches carry large
uncertainties currently. It seems that a measurement with higher
accuracy needs to be done in order to further well constrain the
density dependence of the nuclear matter symmetry energy
coefficient.

\section{Summary}\label{intro}\noindent
We have derived the density functional of the symmetry energy for
finite nuclei in the framework of the Skyrme-Hartree-Fock approach.
The nuclear symmetry energy coefficient $a_{\text{sym}}({A})$ has
been then directly extracted and investigated. The ratio of the
surface symmetry coefficient to the volume symmetry coefficient
$\kappa$ was also estimated. For $^{208}$Pb, $a_{\text{sym}}({A})$
was calculated to be 20.8-23.6 MeV with the Skyrme interactions
SIII, SLy4, SLy5 and SkM*, and the corresponding ratio $\kappa$ is
$1.8-2.6$. For a heavy nucleus, its surface region contributes
largely to the symmetry energy according to the spatial distribution
of the symmetry energy. It has been shown that the contribution from
the gradient term to the symmetry energy is found to be rather small
(less than $2\%$ for $^{208}$Pb) and the second order term in the
kinetic energy part of the symmetry energy density is also less
important for the symmetry energy (its contribution is less than
$3\%$ for $^{208}$Pb). Meanwhile, an alternative method (a
macroscopic method) has been developed in the present study to
determine the symmetry energy coefficient of heavy nuclei which
resorts to the available experimental alpha decay energies of heavy
nuclei. The value of $\kappa$ was calculated to be
$2.21^{+0.56}_{-0.55}$ with $S_{0}=31.6\pm2.2$ MeV based on the
macroscopic method, which agrees well with that from the Skyrme
density functional approach. Moreover, the two completely different
approaches may validate each other to achieve more compelling
results. The calculated $a_{\text{sym}}({A})$ of $^{208}$Pb with the
macroscopic method was furthermore used to analyze the density
dependence of the symmetry energy coefficient of nuclear matter and
evaluate the values of the slope and curvature parameters $L$ and
$K_\text{{sym}}$. It was found that $L=61^{+22}_{-22}$ MeV and
$K_\text{{sym}}=-109^{+71}_{-70}$ MeV. The neutron skin thickness of
$^{208}\text{Pb}$ was then estimated to be $\Delta
R_{np}=0.191\pm0.032$ fm, which is consistent with the pygmy dipole
resonance studies.

This work was supported by the Major State Basic Research Developing
Program of China under No. 2013CB834405, the National Natural
Science Foundation of China under Grants No. 11175219, 10975190, and
11275271; the Knowledge Innovation Project (KJCX2-EW-N01) of Chinese
Academy of Sciences, CAS/SAFEA International Partnership Program for
Creative Research Teams (CXTD-J2005-1)and the Funds for Creative
Research Groups of China under Grant No. 11021504.

\end{CJK*}

\end{document}